\documentclass[mathleft
]{an}
\usepackage{graphicx,natbib}
\usepackage{times}
\begin{document}

\Pagespan{789}{}
\Yearpublication{2006}%
\Yearsubmission{2005}%
\Month{11}%
\Volume{999}%
\Issue{88}%

\title{XMM-Newton Publication Statistics}

\author{J.-U. Ness\inst{1}\fnmsep\thanks{Corresponding author:
  \email{juness@sciops.esa.int}\newline}
 \and A.N. Parmar\inst{2} \and
    L.A. Valencic\inst{3,4} \and R. Smith\inst{5} \and
    N. Loiseau\inst{1} \and A. Salama\inst{1}\thanks{A.S. sadly deceased before the project was completed} \and
    M. Ehle\inst{1} \and N. Schartel\inst{6}
}
\titlerunning{XMM-Newton Publication Statistics}
\authorrunning{J.-U. Ness et al.}
   \institute{XMM-Newton Science Operations Centre, Science Operations
        Department of ESA, ESAC, 28691 Villanueva de la Ca\~nada (Madrid), Spain;
           \email{juness@sciops.esa.int}
        \and
       Science Support Office, Directorate of Science and Robotic Exploration,
       ESA/ESTEC, Postbus 299, NL-2200 AG Noordwijk, The Netherlands
        \and
        Johns Hopkins University, Baltimore, MD 21218
        \and
        NASA Goddard Space Flight Center, Greenbelt, MD 20771
        \and
        Harvard Smithsonian Center for Astrophysics, 60 Garden Street, Cambridge, MA 02138 USA
        \and
        Astrophysics and Fundamental Physics Missions Division, Research and Scientific Support
       Department of ESA, ESAC, Villanueva de la Ca\~nada (Madrid), Spain
             }

\received{31 July 2013}
\accepted{}
\publonline{later}


\abstract{%
We assessed the scientific productivity of XMM-Newton by examining publications and data usage statistics.
We analyse 3272 refereed papers, published until the end of 2012, that directly use XMM-Newton data. The SAO/NASA Astrophysics Data System (ADS) was used to provide additional information on each paper including the number of citations. For each paper, the XMM-Newton observation identifiers and instruments used to provide the scientific results were determined. The identifiers were used to access the XMM-Newton Science Archive (XSA) to provide detailed information on the observations themselves and on the original proposals. The information obtained from these sources was then combined to allow the scientific productivity of the mission to be assessed.
Since around three years after the launch of XMM-Newton there have been around 300 refereed papers per year that directly use XMM-Newton data. After more than 13 years in operation, this rate shows no evidence that it is decreasing. Since 2002, around 100 scientists per year became lead authors for the first time on a refereed paper which directly uses XMM-Newton data. Each refereed XMM-Newton paper receives on average around four citations per year in the first few years with a long-term citation rate of three citations per year, more than five years after publication. About half of the articles citing XMM-Newton articles are not primarily X-ray observational papers. The distribution of elapsed time between observations taken under the Guest Observer programme and first article peaks at 2 years with a possible second peak at 3.25 years. Observations taken under the Target of Opportunity programme are published significantly faster, after one year on average. The fraction of science time taken until the end of 2009 that has been used in at least one article is $\sim 90$\%. Most observations were used more than once, yielding on average a factor of two in usage on available observing time per year. About 20\% of all slew observations have been used in publications.
The scientific productivity of XMM-Newton measured by the publication rate, number of new authors and citation rate, remains extremely high with no evidence that it is decreasing after more than 13 years of operations.
}
\maketitle

\section{Introduction}

XMM-Newton was launched on 10 December 1999 into a 48-hour highly elliptical orbit.
The mission provides sensitive X-ray imaging and spectroscopic observations
of a wide variety of cosmic sources from nearby solar system objects to the most distant
black holes \citep{jansen01}. The payload consists of the European
Photon Imaging Camera
(EPIC), the Reflection Grating Spectrometer
(RGS; \citealt{rgs}) and the Optical Monitor (OM; \citealt{om1}).
The EPIC consists of 3 imaging spectrometers each located at the focus of an X-ray optic consisting of 58 nested Wolter
I geometry mirrors. Two of the EPIC cameras are based on MOS-CCD technology and share the mirrors with RGS grating arrays
\citep{epic_mos} while the detector based on pn-CCD technology is located behind a fully open telescope
\citep{epic_pn}. The overall effective aperture is 4500 cm$^2$ at 1~keV and the spatial resolution is 15 arc seconds
(half-energy width) with a field of view of $\sim$30 arc minutes diameter. The RGS provides 0.35--2.4~keV spectra
with an E/$\Delta$E of 300--700 (1$^{\rm st}$ order).
The effective area for the two grating arrays varies in the range of 40--200~cm$^2$ over the energy range.
The OM provides optical and UV monitoring of fluxes through various filters as well as spectroscopy with two grisms.
Normally, all three instruments are operated simultaneously.\\

XMM-Newton observing time is allocated in different ways:
\begin{itemize}
\item The majority of the observing time is available to the
world-wide scientific community in the Guest Observer (GO) programme.
Targets are selected competitively through peer review by a Time Allocation
Committee which evaluates proposals. The calls
for proposals are normally issued annually and are typically six times oversubscribed in terms of available
observing time. Investigator teams of successful proposals are granted
a proprietary period of one year before the data are made publicly
available.
\item During the first two years of the mission, the scientific groups that
were involved in the instrument and scientific ground segment development
were awarded Guaranteed Time observations.
\item XMM-Newton undertakes
a small number of calibration observations which are made immediately accessible to the
public.
\item Target of Opportunity (ToO) observations
allow XMM-Newton to respond to unique events that sometimes require
quick reaction. Members of the scientific community can alert the
Project Scientist who may approve observations after evaluation.
\end{itemize}

The scientific products from each observation, produced by a
pipeline processing, are stored in the XMM-Newton Science Archive (XSA).
During any proprietary period, data in the XSA are only made available to
the Principal Investigator (PI) of the observation, otherwise all the products
stored in the XSA are available for public download via the internet.
The XSA has more than 3000 registered users of which around 1000 download
data each year. General information such as observing date and
time, instrument use, name of PIs
etc. can be accessed by the general public for all observations
in the XSA.
The key to access a data set belonging to a specific observation
is the unique 10-digit observation identifier (ObsID).\\

Here we investigate the scientific productivity of the XMM-Newton
observatory. We have linked the papers to the XMM-Newton observations
that were used to provide the data that were analysed. XMM-Newton
has a large number of publications
(currently over 3500 refereed papers) which have resulted from data
obtained over more than 13 years of observations and published by
more than 1200 first authors. In the following, we describe the
criteria by which publications are included in the XMM-Newton
publication list, the process by which a publication database
is populated with the necessary information such as the
bibliographic codes, instruments and observations that were used in
each paper. We then use this publication database and the links to
the data to investigate the scientific productivity of XMM-Newton.\\

\section{Generation of the XMM-Newton publication database}

   \begin{table}
   \begin{center}
    \caption{Publication classification scheme used in this paper
    (following \citealt{apia2010} and \citealt{chandrapubs})}
    \begin{tabular}{ll}
      \hline\hline\noalign{\smallskip}
      Category& Description \\
      \noalign{\smallskip\hrule\smallskip}
1 &The publication makes direct use of XMM-Newton\\
      & data \\
  2 & The publication refers to published results \\
  3 & The publication predicts XMM-Newton results \\
  4 & The publication describes XMM-Newton \\
      & instrumentation, software, or operations \\
  5 & Other XMM-Newton related articles\\
        \noalign{\smallskip\hrule\smallskip}
      \label{tab:classification}
    \end{tabular}
   \end{center}
   \end{table}

\subsection{XMM-Newton publication list}

The first step is to define the papers that are to be considered
as XMM-Newton papers for this study.
\cite{apia2010} and \cite{chandrapubs} have defined five categories
of mission-related publications which are listed in
Table~\ref{tab:classification}.\\

Since the start of the XMM-Newton mission, the XMM-Newton Guest Observer
Facility, at the Goddard Space Flight Center (GSFC), has kept a
detailed record of the bibcodes of {\it all} refereed XMM-Newton
publications\footnote{http://heasarc.gsfc.nasa.gov/docs/xmm/xmmbib.html}.
Published articles are considered as XMM-Newton papers if they are
published in refereed journals, mention ``XMM-Newton'' or ``X-ray''
in the abstract, and include XMM-Newton data or results.
This can be done either through directly reducing and analysing
XMM-Newton data, discussing in detail previous XMM-Newton results,
using data obtained from XMM-Newton in plots or text, or using
XMM-Newton survey results.\\

We include in this study
only category I papers from Table~\ref{tab:classification},
those that make direct use of XMM-Newton data including X-ray
fluxes or hardness ratios from the 2XMM catalogue
\citep{watson09} and the slew survey \citep{saxton08}.
We also add papers from the special issue of Astronomy
\& Astrophysics (vol. 365) in 2001 which included the
descriptions of the mission, instruments and ground segment
(category 4 in Table~\ref{tab:classification}).\\

The GSFC archive of XMM-Newton papers is regularly updated with
newly published papers. Since the GSFC list includes all the
publication categories listed in Table~\ref{tab:classification}, the
XMM-Newton Project Scientist at ESA manually checks each listed paper
to select those that fulfil the criteria discussed above. The list
of papers derived in this way is then used to provide information
to ESA management and elsewhere.
As of autumn 2013, there are more than 3500 refereed papers that
fulfil the criteria listed above out of which we analyse
3272 that were published until the end of 2012.

\subsection{Observation identifier (ObsID) determination}

The next step is to derive the XMM-Newton observation identifiers (ObsIDs)
for the observations that were analysed to provide the scientific
results in each paper. These, and the instruments used, were extracted
manually from the papers
with the help of the other scientists given in acknowledgements.
Whilst the XMM-Newton publication
guidelines\footnote{http://xmm.esac.esa.int/external/xmm\_science/pub\_guide.shtml}
request that the ObsID(s) and instruments be included in the
article, only about 40\% of papers actually contain this information.
Observations
that are not used directly for scientific purposes but, e.g.,
for determination of instrumental background are not included
in the database.
For the majority of papers which did not provide ObsIDs this
was still a relatively straight forward process, utilising
information such as the target name, coordinates, observation
date, or instrumental setup if provided.
Unique catalogue names from the 2XMM catalog \citep{watson09} and the
slew survey \citep{saxton08}, if given by the authors,
allow extraction of the corresponding ObsIDs from the catalog.\\
Care was taken with
papers that used large numbers of observations such as sky surveys,
for which observations have been selected by the authors from criteria
that are not uniquely reproducible. With the co-operation of the
corresponding authors, most of these papers (38) could be
resolved, although some uncertainty of the order 5\% remains.
We flagged 29 difficult papers for which the ObsIDs included in
the database are particularly uncertain.
To minimise human errors, each article was screened independently
by scientists in the XMM-Newton Guest Observer Facility at the
Goddard Space Flight Center (GSFC), USA and in the XMM-Newton
Science Operations Centre (SOC) in Spain. The results from both
groups were then compared and discrepancies resolved.\\

   \begin{table}
      \caption{\label{tab:dbdiagram}Contents of each entry in the XMM-Newton
publication database}
      \begin{tabular}{l|r}
\hline
Source & Item \\
\hline
NASA GSFC & Publication bibcode \\
\hline
ESA SOC & ObsID(s) of data analysed in paper \\
 \& GSFC                & Instruments used to provide analysed data \\
\multicolumn{2}{l}{{\bf Publication database} used in this work}\\
\multicolumn{2}{l}{\ \ \  3272$^a$ Papers analysed in this work}\\
\multicolumn{2}{l}{\ \ \  3219$^b$ Papers with ObsIDs (98.4\% of previous)}\\
\multicolumn{2}{l}{\ \ \  3187$^b$ Papers with $<290$ ObsIDs (99.0\% of previous)}\\
\multicolumn{2}{l}{\ \ \  79348 ObsIDs in total}\\
\multicolumn{2}{l}{\ \ \  72634$^b$ certain ObsIDs (91.5\% of previous)}\\
\multicolumn{2}{l}{\ \ \  7131 unique ObsIDs}\\
\multicolumn{2}{l}{\ \ \  7096$^b$ certain unique ObsIDs (99.5\% of previous)}\\
\\
\hline
{\bf ADS}    & Publication Date\\
           & Journal name \\
           & Authors name(s) \\
           & Paper title and abstract \\
           & bibcode(s) of citing paper(s) \\
\multicolumn{2}{l}{{\bf Citation database} of refereed citing articles}\\
\multicolumn{2}{l}{\ \ \  published until 10 October 2013}\\
\multicolumn{2}{l}{\ \ \ 74564$^a$ Refereed citing papers}\\
\multicolumn{2}{l}{\ \ \ 20295$^a$ unique citing papers}\\
\\
\hline
{\bf XSA}    &  Observation date \\
           &  Exposure time \\
           &  Name(s) of proposers \\
           & Proposed science category \\
\multicolumn{2}{l}{\ \ \ 8577 usable observations (see Sect.~\ref{sect:analysis})}\\
\hline
      \end{tabular}

$^a$Used in Sect.~\ref{sect:primres}\\
$^b$Used in Sect.~\ref{sect:secres}\\
The motivation for the sub-samples is outlined in Sect.~\ref{sect:results}
     \end{table}

\subsection{Publication database}

Information for each XMM-Newton article is stored in a
database whose contents are given in Table~\ref{tab:dbdiagram}.
The bibcodes allow the ADS to be accessed and the information listed
under ADS in Table~\ref{tab:dbdiagram} to be obtained. Similarly the
ObsIDs allow the XSA to be accessed and the information listed under
XSA to be extracted. In addition, the submission dates were extracted
directly from the articles via automatic keyword search ($\sim$95\%)
and manual checks ($\sim$2\%). The maximum delay between submission
and publication dates listed in the ADS is 4.7 years whilst on
average, it is six months.
The ADS also allows cross links between publications that make
reference to each other which were used to create a citation database
of {\em refereed} articles citing XMM-Newton articles. We analyse only
papers published until the end of 2012 whilst we include citations
that were included in the ADS until 10 October 2013. The citation
database was populated by extracting for each article in the
publication database a list of bibcodes from the ADS that belong to
refereed articles referring to it.\\

The contents of the ADS originate from metadata
provided by the journals which in turn receive the information from
the authors when submitting the papers. References may be incomplete due to
the inability to match them with 100\% accuracy (e.g. in press, private
communications, author errors, etc.). The robustness of citations in the
ADS was compared to the Science Citation Index (SCI) by H. A. Abt
(published in
\citealt{heck06}\footnote{http://www.garfield.library.upenn.edu/papers/...\\
...helmutabtorgstratastronv6y2004.html})
and demonstrates the power of ADS with 15\% more citations than SCI
and missing less than 1\% of the citations.
Some caveats when analysing citations as criterion to assess the
impact of an article are described, e.g., in \cite{chandrapubs}, and
only rough measures are possible to deal with contaminating effects
such as self citations. However, a systematic study
presented by \cite{trimble86} analysing articles that were
published during January 1983 revealed that only about 15\% of all
citations were actually self citations and that the rate varies very
little (8\%) among 33 world wide journals. In light of
the complexity of any corrective measure and reproducibility of
results, we use the citation statistics as they are extracted from
the ADS.\\

\section{Analysis}
\label{sect:analysis}

The bibcodes and ObsIDs allow access to
the ADS and the XSA to provide details about the
papers and the observations, respectively.
In Sect.~\ref{sect:primres}, we present results
that rely only on information extracted from the ADS which
allows items such as publication rate, journal statistics,
author statistics, and citation statistics to be investigated.
Although the ADS operates a sophisticated system that streamlines
author names, the identification of individual authors is not
straightforward. We identified individuals by their full last
name and first initial. This will count
different authors with common last names such as John Smith
and Jim Smith as a single author which is a conservative
approach for our purposes. On the other hand, we avoid
the larger number of same authors with different entries of
first name such as J. Smith, J.H. Smith, John H. Smith,
J. Hamilton Smith, etc. to be counted multiple times.
Double counts can also arise from different spellings of
last names, in particular if they contain special characters
such as Spanish or French accents or German umlauts.
We perform systematic replacements
of all special characters in the ADS lists of authors by utf-8
compliant characters to minimise the double counting of different
authors.\\

To assess the usage of XMM-Newton data presented in
Sect.~\ref{sect:secres}, we determined the elapsed time between the
observation and
first publication of the data, the fraction of the observing time used in
publications, multiple use of the same XMM-Newton observation and usage
of data from different science programmes and science categories.
The data for these studies originates from the combined
information obtained by linking the bibcodes and ObsIDs.
This linking allows publication dates and
citation statistics obtained from the ADS to be cross-correlated with
observing dates, exposure times, science programmes and categories
etc. obtained from the XSA.\\

The reference sample of useable observations,
suitable for scientific analysis, comprises 8577 pointed
observations, obtained until the end of 2012. The definition
of "useable" is that the automatic generation of the Observation
Data File (ODF) was successful and the products from the
Pipeline Processing System (PPS) are availale from the current
version of the XSA.
Observations with only OM exposures are excluded from
the reference sample. We added 94 observations for which
no PPS products had been released, but where at least one
EPIC camera was operated in a science mode with more
than 1000\,seconds exposure time. We did not screen
the quality reports, nor did we filter on good time
intervals. The reference sample is thus a rough
estimate of observations that are available
to the scientific community and is used to investigate
the degree to which the science products were used.\\

As a measure of the impact of an observation
in the reference sample, we determined the number
of papers using it, the date when the first
paper was published,
and the number of refereed papers citing all articles
combined using the particular observation.\\

\section{Results}
\label{sect:results}

\begin{figure*}[!ht]
\resizebox{\hsize}{!}{\includegraphics{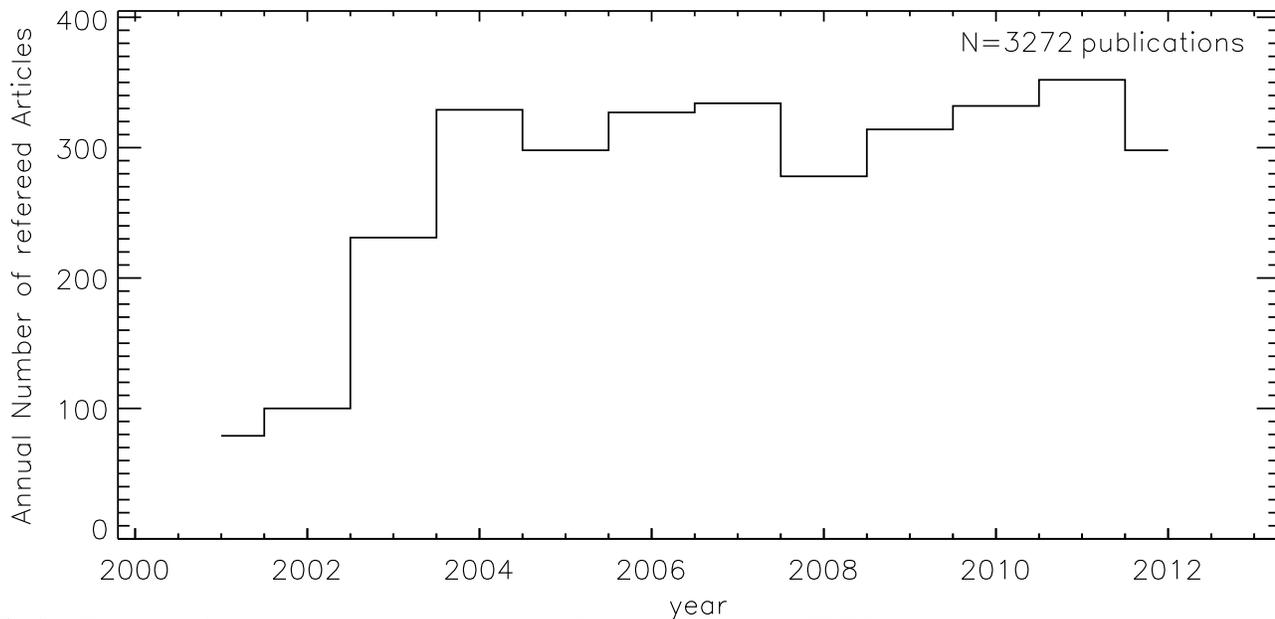}}
\caption{Evolution of the annual number of published refereed
papers that use XMM-Newton data, based on the publication dates
extracted from ADS. Since 2003, around 300 new XMM-Newton
papers are published every year.
}
\label{fig:publications}
\end{figure*}

The publication database up to the end of 2012 contains
in total 3272 articles, out of which we were able to identify
ObsIDs for 3219. 29 articles with uncertain lists of ObsIDs
were discarded. The citation database contains all
refereed articles that were published until 10 October 2013.
In Table~\ref{tab:dbdiagram}, we give an account of the
contents of the publication and citation databases with
total number of entries. Those samples are used in subsections
\ref{sect:primres}
and \ref{sect:secres} are marked by superscripts $a$ and
$b$, respectively.\\
%


\subsection{Publication statistics}
\label{sect:primres}

\subsubsection{Publication rates}

Based on the publication dates extracted from the ADS,
we determined the number of published papers per calendar
year and show the evolution in Fig.~\ref{fig:publications}.
Following launch in December 1999, the first
refereed papers appeared in 2001. By 2004-2005, a stable
publication rate of around 300 papers per year had been reached.
It is remarkable that the publication rate shows no evidence
for any reduction, even more than 13 years after launch.

\subsubsection{Publication journals}
\label{sec:journals}

   \begin{table*}
   \begin{center}
      \caption{Nearly 96\% of all
      XMM-Newton refereed papers are published in the nine journals
      listed. The journal
      abbreviations used in the bibcodes are given in brackets.
        }
      \begin{tabular}{lcccc}
      \hline\hline\noalign{\smallskip}
      Journal & Number of& \% & Number of & \% \\
      & publications& & citations$^a$ \\
      \noalign{\smallskip\hrule\smallskip}
Astronomy \& Astrophysics (A\&A) & 1170 & 35.8 & 4002 & 19.7\\
Astrophysical Journal (ApJ) & 1015 & 31.0 & 7478 & 36.8\\
Monthly Notices of the Royal & 659 & 20.1 & 4371 & 21.5\\
\ \ Astronomical Society (MNRAS) &&&&\\
Astronomical Journal (AJ) & 66 & 2.02 & 616 & 3.04\\
Publications of the Astronomical & 48 & 1.47&400&1.97\\
\ \ Society of Japan (PASJ) &&&&\\
Astronomische Nachrichten (AN) & 77 & 2.35 & 241& 1.19\\
Advances in Space Research (AdSpR) & 72 & 2.20 & 180&0.89 \\
Nature + Science & 24 & 0.73 & 229& 1.13\\
\hline
\hfill Total:&3131&95.7&17517&86.3\\

       \noalign{\smallskip\hrule\smallskip}
      \label{tab:journals}
      \end{tabular}

$^a$as of 10 October 2013
      \end{center}
     \end{table*}

The most popular journal in which to publish XMM-Newton papers is
Astronomy \& Astrophysics (A\&A) followed by the Astrophysical Journal
(ApJ), and Monthly Notices of the Royal Astronomical Society (MNRAS);
87\% of all XMM-Newton papers have been published in
one of these three journals. Extending this to the nine journals
listed in Table~\ref{tab:journals} allows almost 96\% of all XMM-Newton
refereed papers to be included.
The last two columns in Table~\ref{tab:journals} give the number and
percentage of papers in each journal that provide
references to the XMM-Newton papers. Here, papers citing more
than one XMM-Newton papers are counted only once.
Whilst the majority of XMM-Newton papers
are published in A\&A, papers appearing in ApJ make the highest
number of references to them.

\subsubsection{Authors}

\begin{figure*}[!ht]
\resizebox{\hsize}{!}{\includegraphics{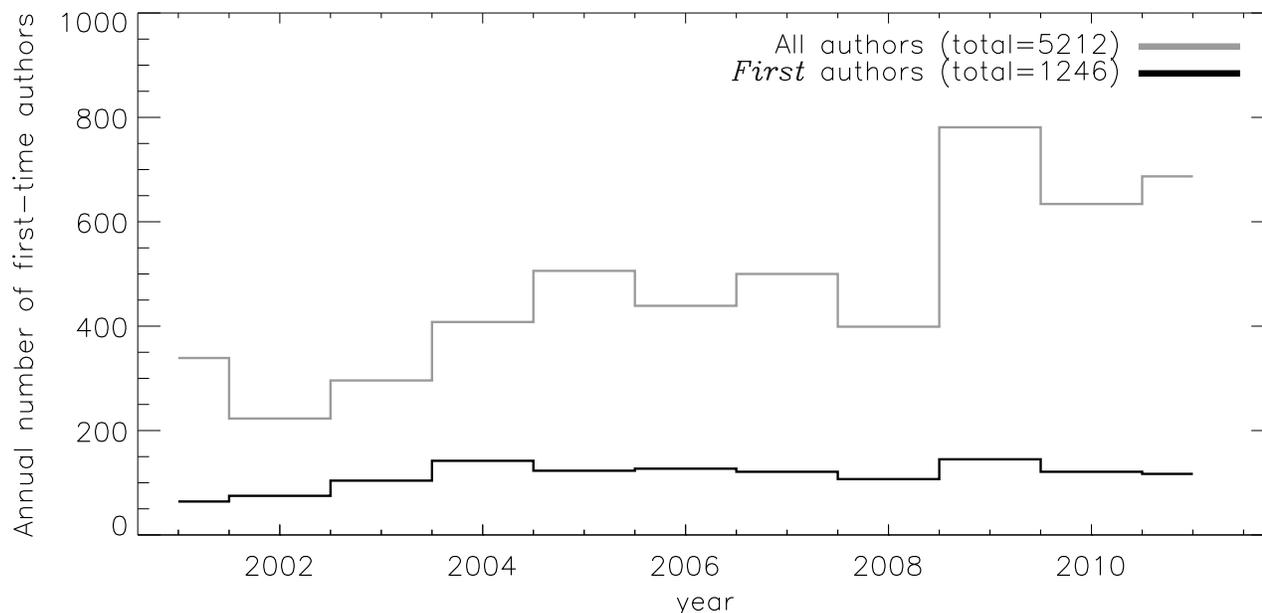}}
\caption{Evolution with time of new authors (grey) and new
{\em first} authors (black) publishing a refereed paper which
directly uses XMM-Newton data.
}
\label{fig:authors}
\end{figure*}

An analysis of first- and co-authors publication rates gives
insights into the XMM-Newton user community. For each unique name
we determined the dates when he or she first appeared as a
first- or co-author of a refereed XMM-Newton paper. The results
are shown in Fig.~\ref{fig:authors} where the number of first
and co-authors publishing an XMM-Newton data paper for the first
time is shown. The grey histogram indicates all authors, whilst
the black histogram counts only first authors. It can be seen
that the number of new first authors has remained approximately
constant at around 100 scientists each year. The number
of new authors may be showing a gradual increase with time.
The fact that there are around 100 new first authors each year
demonstrates that the XMM-Newton user community continues to grow,
demonstrating the continued relevance of the mission to the
wider scientific community.\\

\subsubsection{Citations}
\label{sect:citations}

\begin{table*}
  \caption{XMM-Newton Papers with highest citation rates until 10 October 2013}
  \begin{tabular}{lllll}
  \hline
bibcode & Authors & Instruments & \#citations & citations per year\\
\hline
2001A\&A...365L..18S & L. Str\"uder et al. & EPN & 1089 & 84.68\\
&\multicolumn{4}{l}{\it The European Photon Imaging Camera on XMM-Newton}\\
2001A\&A...365L..27T & M. J. L. Turner et al. & EMOS & 1013 & 78.77\\
&\multicolumn{4}{l}{\it The European Photon Imaging Camera on XMM-Newton}\\
2001A\&A...365L...1J & F. Jansen et al. & XMM & 750 & 58.32\\
&\multicolumn{4}{l}{\it XMM-Newton observatory. I. The spacecraft and operations}\\
2010A\&A...517A..92A & M. Arnaud et al. & EPN, EMOS & 157 & 46.72\\
&\multicolumn{4}{l}{\it The universal galaxy cluster pressure profile from a representative}\\
&\multicolumn{4}{l}{\it sample of nearby systems (REXCESS) and the YSZ - M500 relation}\\
2009A\&A...493..339W & M. G. Watson et al. & EPN, EMOS & 202 & 41.56\\
&\multicolumn{4}{l}{\it The XMM-Newton serendipitous survey. V.}\\
&\multicolumn{4}{l}{\it The Second XMM-Newton serendipitous source catalogue}\\
2005A\&A...441..417H & G. Hasinger et al. & EPN, EMOS & 303 & 37.36\\
&\multicolumn{4}{l}{\it Luminosity-dependent evolution of soft X-ray selected AGN.}\\
&\multicolumn{4}{l}{\it New Chandra and XMM-Newton surveys}\\
2007ApJ...663...81P & M. Polletta et al. & EPN, EMOS & 233 & 36.63\\
&\multicolumn{4}{l}{\it Spectral Energy Distributions of Hard X-Ray Selected Active}\\
&\multicolumn{4}{l}{\it Galactic Nuclei in the XMM-Newton Medium Deep Survey}\\
2001A\&A...365L...7D & J. W. den Herder et al. & RGS & 421 & 32.74\\
&\multicolumn{4}{l}{\it The Reflection Grating Spectrometer on board XMM-Newton}\\
2009A\&A...498..361P & G. W. Pratt et al. & EPN, EMOS & 143 & 31.59\\
&\multicolumn{4}{l}{\it Galaxy cluster X-ray luminosity scaling relations}\\
&\multicolumn{4}{l}{\it from a representative local sample (REXCESS)}\\
2003ApJ...590..207P & J. R. Peterson et al. & RGS & 306 & 29.30\\
&\multicolumn{4}{l}{\it High-Resolution X-Ray Spectroscopic Constraints}\\
&\multicolumn{4}{l}{\it on Cooling-Flow Models for Clusters of Galaxies}\\

  \hline\hline
\label{tab:citbest}
  \end{tabular}
\end{table*}

In Table~\ref{tab:citbest}, we list the XMM-Newton papers with the
highest citation rates, giving bibcodes, first author, instruments
used, number of citations, annual citation rate and titles.
Only papers with an exposure to the public of more than two
years were considered (thus were published until the end of 2010)
in order to eliminate short-term effects. The highest citation
rates were received by the instrument and 2XMM catalog papers.\\

\begin{figure}[!ht]
\resizebox{\hsize}{!}{\includegraphics{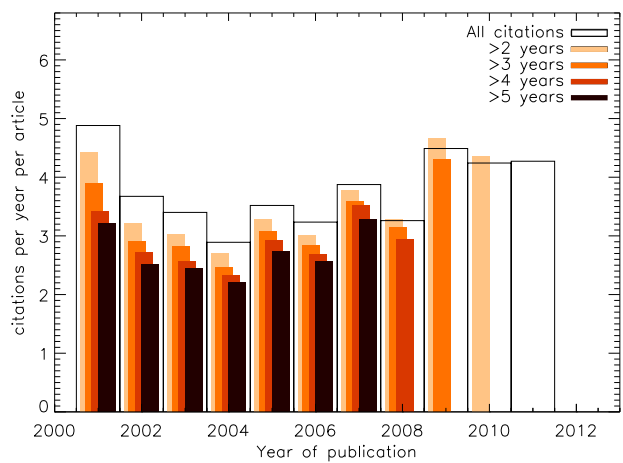}}
\caption{Combined annual citation rates received by all papers
within each year, normalised by the corresponding number of papers
in the same time interval. The shaded bars indicate the citation rates
by papers that appeared more than 2, 3, 4, and 5 years after the
cited papers were published to illustrate the long-term impact.
}
\label{fig:citevol}
\end{figure}

In Fig.~\ref{fig:citevol}, we show the number of citations per year
per number of papers published in each corresponding
year (=number of citable items). To assess the longer-term impact of
the mission we have also determined the fraction of citations that were given
more than 2, 3, 4, and 5 years after publication and mark these with
different shades of red in Fig.~\ref{fig:citevol}
(see legend). The reference time period used to determine the rate
is the time between publication and 10 October 2013. For the long-term citation
rates, the reference period is accordingly shorter. For example, for
a paper published in January 2003, the total reference period is
10 years, whilst the reference period for citations given later than
3 years after publication is only 7 years.
In general, papers published shortly after the launch of the mission
have received the highest number of citations (note that citations to
instrument articles \cite{epic_pn, epic_mos,rgs,om1,om,jansen01} are excluded
here). After 2002, the number of citations has remained at an average rate of
3-4 citations per year per paper, e.g., a paper published in 2003 has on average
been cited 38 times. The long-term citation rates for all four
reference periods remain remarkably high compared to the shorter-term
citation rate. We emphasise that the corresponding citation rates
include citing papers from the most recent past which shows the
continuing relevance of XMM-Newton results.\\

\begin{figure}[!ht]
\resizebox{\hsize}{!}{\includegraphics{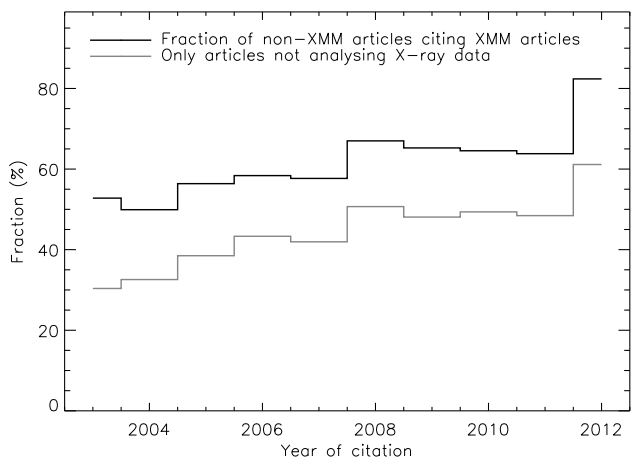}}
\caption{The fraction of citing papers that do not directly use
XMM-Newton data
(black histogram). In addition, the same fraction is determined
for papers that do not
primarily discuss the analysis of X-ray data.
}
\label{fig:beyond}
\end{figure}

In order to estimate the impact of XMM-Newton beyond the X-ray
community, we determined the fraction of papers that cite
XMM-Newton papers that are not included in our publication database.
The evolution of this fraction is shown by the black histogram in
Fig.~\ref{fig:beyond}, starting with the year 2003, when the publication
rate has reached a stable level (see Fig.~\ref{fig:publications}).
In addition, we determined the
(lower) fraction of these articles that do not contain the names
of other current major X-ray missions {\it Chandra}, Suzaku, or Swift in their titles
or abstracts and show the respective fractions with the grey histogram.
As more XMM-Newton papers were published, the fraction of non-XMM-Newton papers
citing XMM-Newton papers has increased.
Also, the fraction of citing papers not primarily analysing X-ray
data has increased, having reached over
50\% in 2012. This shows that the high citation rate is not
generated by the X-ray community itself, but that XMM-Newton
results are of interest to a much larger scientific community.\\

\subsection{Usage Statistics}
\label{sect:secres}

In order to put statistics on used observations into a context,
the reference sample of useable observations was defined in
Sect.~\ref{sect:analysis}, comprising 8577 pointed observations.
We exclude slew exposures in the analysis that follows
and instead performed a quick separate assessment: 2738 slew
observations have been
taken up to the end of 2012, 603 of these (22\%) have been used
(excluding the article by \citealt{saxton08}), which is
a high number considering that slew exposures are taken at
random positions and at random times. 251 slew observations
have been used more than once, up to five times, yielding
a total of 926 slew observations that have been used
in scientific articles including multiple use.\\

\begin{figure}[!ht]
\resizebox{\hsize}{!}{\includegraphics{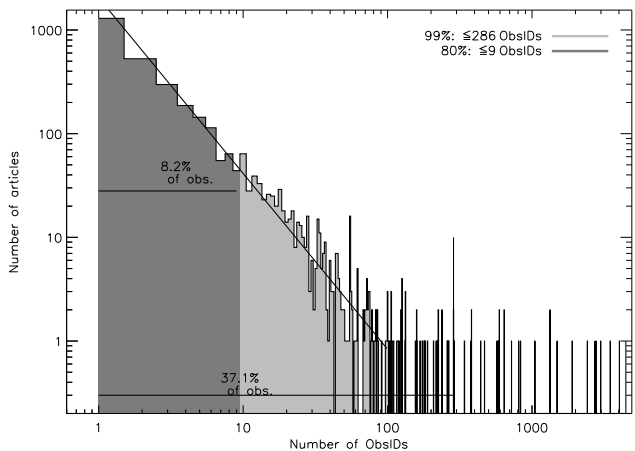}}
\caption{The distribution of the number of observations used per paper
follows an approximate power-law distribution with
the majority of papers using data from less than
three observations. 1\% of XMM-Newton papers
use data from more than 290 different observations. The shades
mark accumulated fractions of all papers, as indicated in
the legend. The vertical bars mark the limits given in the legend,
and the percentages give the fraction of all observations used
in the corresponding sub-samples of articles.
}
\label{fig:obsnum}
\end{figure}

We now present results that rely both on bibliographic
information from the ADS and on the instrument and ObsIDs determined
by manual screening. The total number of papers and ObsIDs in the
publication database are listed in Table~\ref{tab:dbdiagram}.
The distribution of the number of observations used in the papers
is shown in Fig.~\ref{fig:obsnum}. The distribution follows a power
law with an index of -1.7, yielding a median of 2.7 observations per
paper. The majority of papers thus present the analysis of
individual observations. In Fig.~\ref{fig:obsnum}, we highlight
sub-samples of 80\% and 99\% of papers with two shades of grey
which analyse up to 9 and 290 observations, respectively. These
two samples cover 8.1\% and 37.2\% of all observations
in the database. 63\% of observations have been used in
1\% of articles making use of more than 290 ObsIDs each.
In order to show potential differences between dedicated
papers and those papers using large samples of data, some
of the analysis presented in this section has been carried
out separately for the full sample of articles and the
99-\% subsample of articles which use less than 290 observations.\\

\subsubsection{Time delay between observation and first publication}

\begin{figure*}[!ht]
\resizebox{\hsize}{!}{\includegraphics{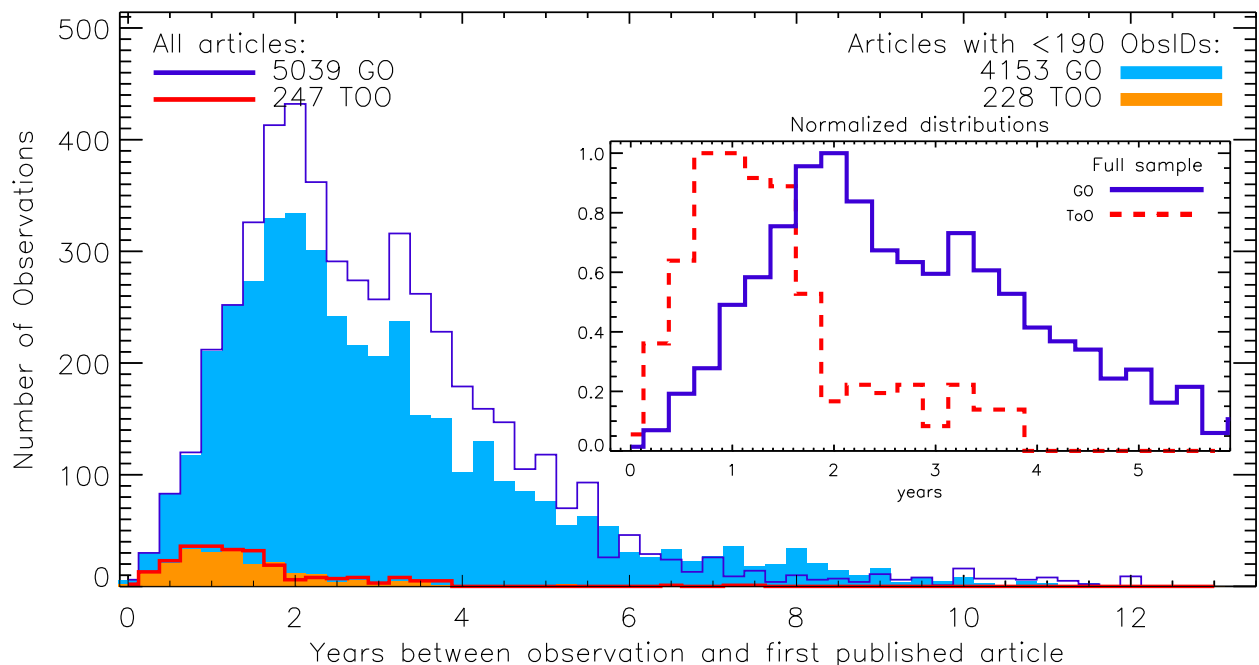}}
\caption{Distribution of elapsed time between observation
and publication dates of the first article using them.
The analysis was performed
for programme categories Guest Observer (GO) and Targets of
Opportunity (ToO), shown with different colours. The 2XMM catalogue
\citep{watson09} is excluded from this analysis.
The shaded distributions represent the results from the
reduced database, excluding 1\% of articles that use more
than 290 ObsIDs. The inset shows the normalised
distribution, illustrating that ToOs are published
faster than regular GO observations.}
\label{fig:firstpubs}
\end{figure*}

In Fig.~\ref{fig:firstpubs}, we show the distribution of elapsed
time between observation and the publication of the first article
for GO and ToO observations. This study was performed for all papers
excluding the 2XMM catalog by \cite{watson09}
(open histograms) and excluding articles using more than 290
observations (99-\% subsample filled histograms). The sample sizes,
i.e., the total number of observations under these two programme
categories used in at least one paper, are given in the legends.\\

The distributions of ToO and GO programmes contain sharp peaks
after approximately 1 and 2 years, respectively.
ToO observations have a proprietary period of 6 months, and the
faster publication of ToO observations reflects this.
The presence of a second peak at 3.25 years
in the distribution of publication times of GO observations is noteworthy.
This occurs one year after the main peak and could reflect
the one year proprietary period before public access to GO data.
The presence of the second peak and the significant
number of papers published many years after the data were
taken indicates that archival research probably play an
important role in the scientific success of XMM-Newton.\\

\subsubsection{Use of available science time}

\begin{figure}[!ht]
\resizebox{\hsize}{!}{\includegraphics{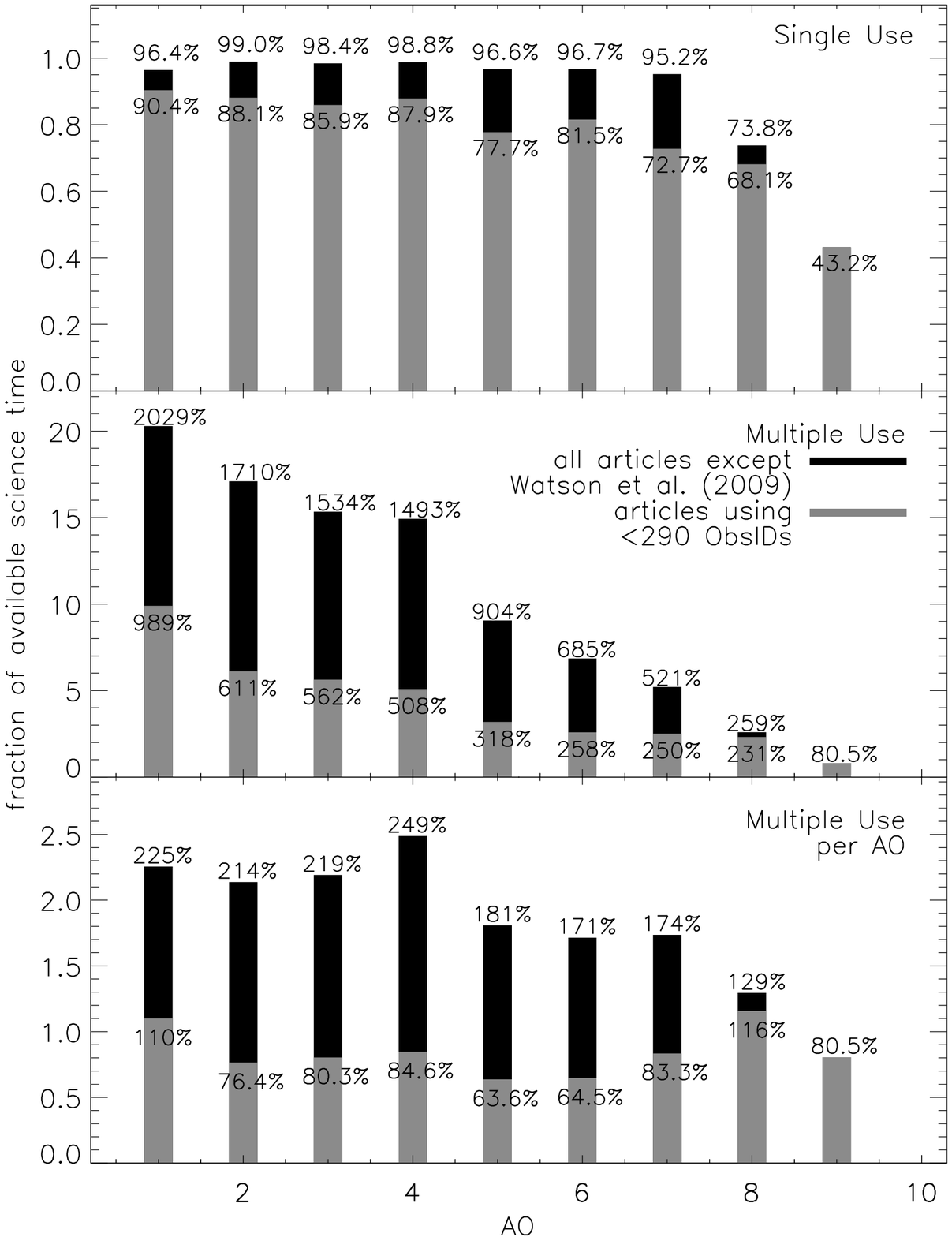}}
\caption{The fraction of available science time used in at least
one publication (top), accounting for multiple use (middle),
and multiple use per observing cycle (AO; bottom). Multiple use is
accounted for by multiplication of the observing time from
each observation by the number of papers using it.
For the black bars, the 2XMM catalogue \citep{watson09}
is excluded while for the grey bars, all papers using more than
290 observations are excluded.
Multiple use naturally increases with the age of the data (middle
panel), yielding a roughly constant recycling {\em rate} (bottom).
}
\label{fig:scitime}
\end{figure}

In Fig.~\ref{fig:scitime}, we present the fraction of
available science time for each observing cycle
(AO)\footnote{Announcement of Opportunity cycles (AOs) last
one year and start on May 1st of each year} that was
used in at least one publication (top panel), in multiple
publications (middle panel), and in multiple publications
per AO (bottom panel) with percentages given on
top of each bar. The 2XMM catalogue by
\cite{watson09} is excluded.
We also computed the same numbers excluding the 1\% of papers
which use more than 290 observations and show these results with
the lighter grey bars. A lower fraction of observations taken in more recent
AOs have been published which is unsurprising given the distribution
of elapsed time between observation and first publication,
illustrated in Fig.~\ref{fig:firstpubs}. More than 95\% of
observing time has been used in scientific publications,
while round 80\% of the time has been used in dedicated papers
focusing on a few observations. The earliest observations
have been recycled up to 20 times, where the fraction
of multiple use scales with age of the data.
The bottom panel shows the rate with which data were re-used
per AO. Each kilosecond of observing time is
used on average twice per year.\\

\begin{figure}[!ht]
\resizebox{\hsize}{!}{\includegraphics{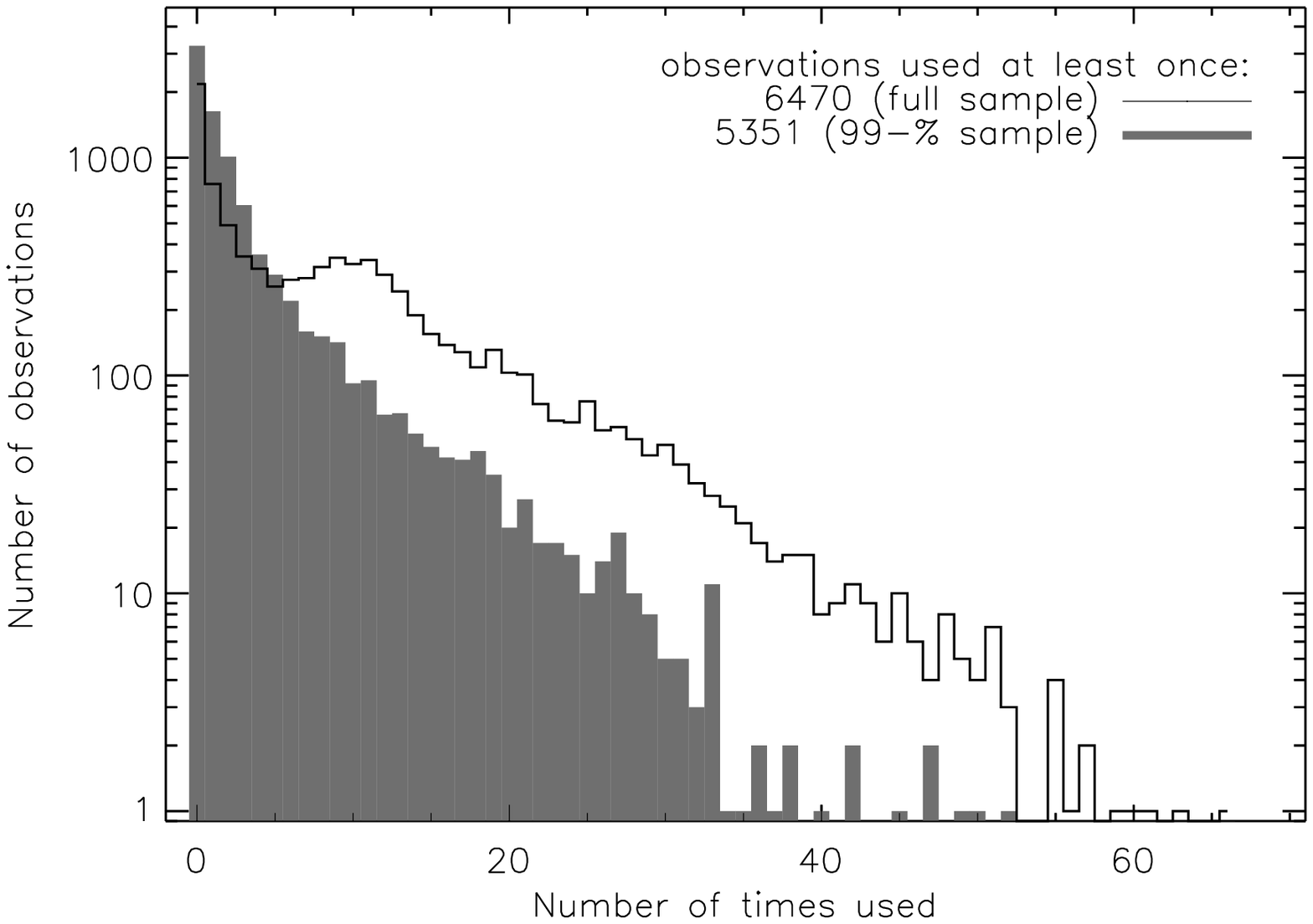}}
\caption{Multiple use of observations used at least once. The
horizontal axis shows the number of papers in which the same
observation was used while the vertical axis counts the number
of observations
that has been used in as many papers given by the horizontal
axis. The grey histogram shows the same distribution for the 99\%
subsample of articles using less than 290 observations.
}
\label{fig:multuse}
\end{figure}

In Fig.~\ref{fig:multuse}, we illustrate
the distribution of multiple usage based on the full publication
database with the open histogram. The observation with the highest
recycling rate is ObsID 0097820101 (the Galaxy Cluster A\,1795)
which has been used in 67 articles.
The exclusion of papers using large numbers of observations
in one project naturally reduces the recycling rate and is shown
in Fig.~\ref{fig:multuse} with the grey histogram. The observation
with the highest recycling rate in the reduced sample is also
ObsID 0097820101 being used 51 times. Also without these
papers, an impressively large number of observations
has been used multiple times.

%

\subsubsection{Science categories}

The majority of XMM-Newton observations were taken in the Guest Observer (GO)
programme. The GO programme is composed of regular observations of
certain targets with observing times up to 300\,ksec,
Large Programs ($>300$\,ksec), and triggered
observations taken in response to anticipated events. In addition,
Target of Opportunity (ToO) observations are performed in response to
unanticipated events. Finally, calibration observations are included at
regular intervals to monitor the performance of each instrument with
data immediately available to the scientific community.\\

\begin{figure}[!ht]
\resizebox{\hsize}{!}{\includegraphics{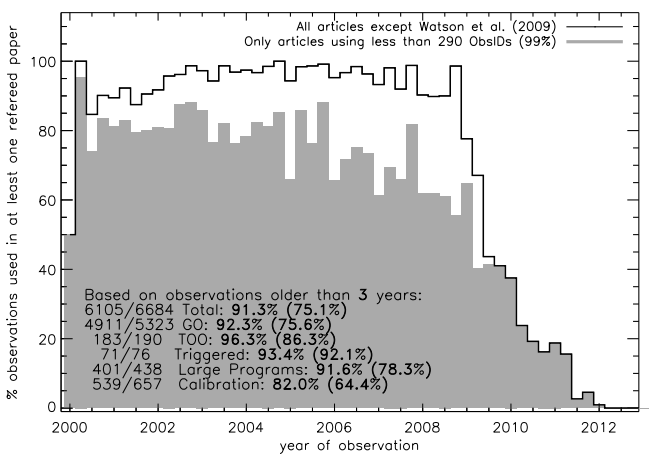}}
\caption{The percentages of observations that
have been used in scientific papers as of end-2012. The black
open histogram accounts for all papers, excluding only observations
used by \cite{watson09} whilst the
light shaded histograms also exclude observations used in
1\% of papers that have analysed more than 290 ObsIDs.
The plot shows the results for regular GO observations
(see text). Average usage for other program categories
is given in the lower left corner, where only observations
older than three years are included in the average.
Given are, for each program, number of observations
used versus number of usable observations, percentage
of total sample and, in brackets, the same fraction resulting
from accounting only for paper from the partial sample of
papers using less than 290 ObsIDs.
}
\label{fig:program}
\end{figure}

In Fig.~\ref{fig:program}, we show the time evolution of the
fraction of available observations that have been used in at
least one article for GO observations (open histogram). Observations
that have been used in the \cite{watson09} are excluded to avoid
overestimation of the usage. The light shaded area indicates the
same results for GO observations having only been used in articles
analysing less than 290 observations. The last three years yield a
lower usage presumably because many observations are still being
analysed (see Fig.~\ref{fig:firstpubs}).

Excluding observations taken within the last three years,
92\% of GO observations have been used to produce a refereed XMM-Newton paper.
About 76\% of all observations have been used in papers which analyse
less than 290 observations. In the bottom left legend of
Fig.~\ref{fig:program}, we give the same average numbers for
the other programmes. 82\% of calibration observations
have been used in all papers combined (excluding only \citealt{watson09})
and 65\% in papers with less than 290 observations, respectively.

\subsubsection{Instrument usage}
\label{sect:insuse}

We studied the instrument usage of all papers except those
describing the instruments themselves. 78\% of them focus on
data from a single instrument, while 19\% use data from two
instruments, and 3.4\% use data from all three instruments.
The majority (96\%) of papers use data from one or both EPIC
detectors. The RGS spectrometer can only be used for bright
point-like sources, naturally leading to less usage (18\%).
Data from the optical monitor (OM) was used in 11\% of all articles.\\

The importance of the results obtained with each instrument can
be measured in terms of the citation rates of the articles
using their data, computed in the same way as in
Sect.~\ref{sect:citations}.\\

\begin{figure}[!ht]
\resizebox{\hsize}{!}{\includegraphics{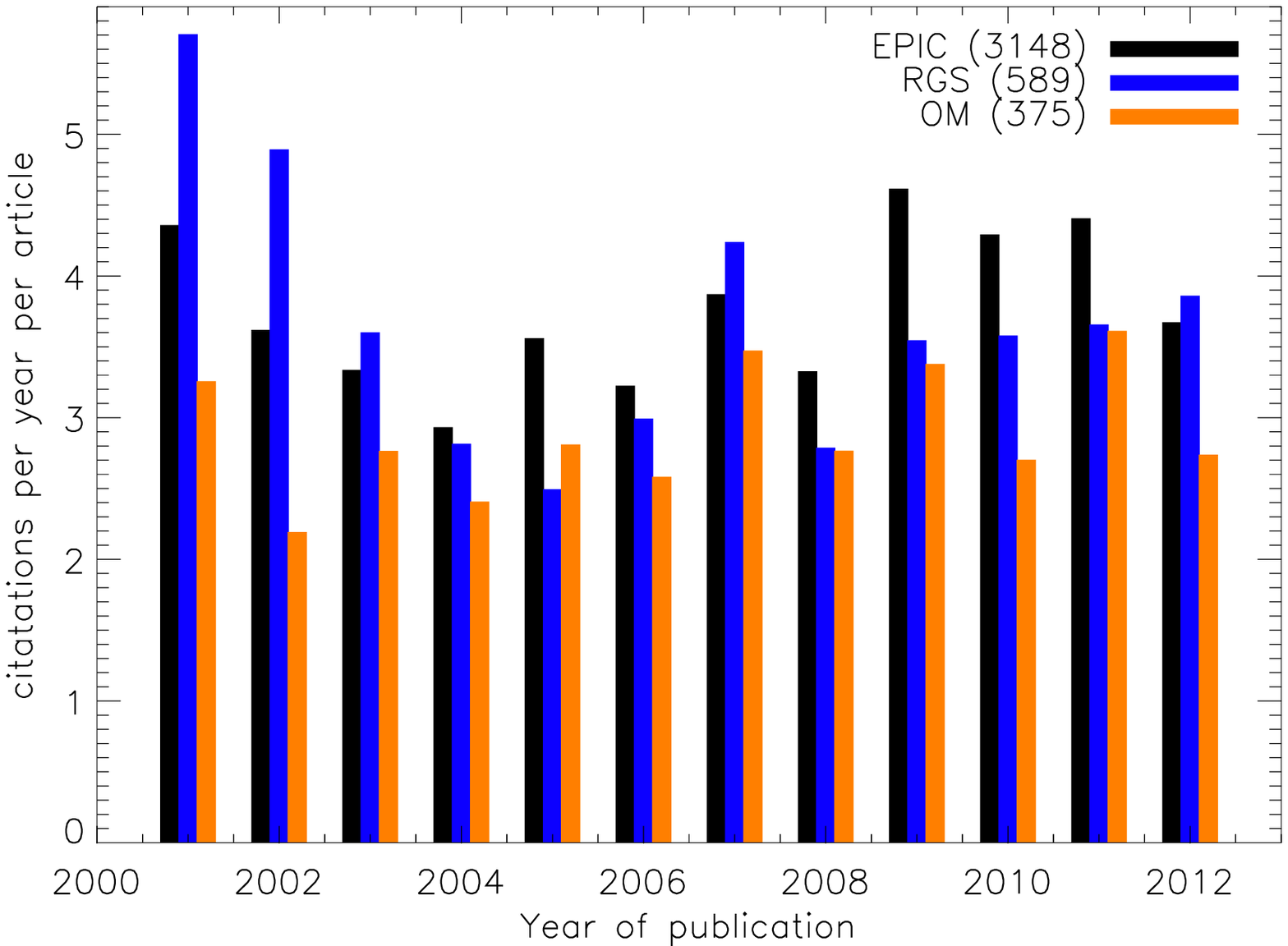}}
\caption{Citation statistics of articles using EPIC,
RGS, and OM data (excluding instrument papers; see legend).
The sample includes 3274 articles, excluding papers
describing the instruments.
Similarly high impact was achieved by all three instruments.
}
\label{fig:ins}
\end{figure}

The evolution of the citation rate for papers analysing
data from different instruments is illustrated in
Fig.~\ref{fig:ins}. The total sample sizes are given in
the right legend and comprise papers using the given
instruments alone or in combination with data from other
instruments.\\

All instruments contribute roughly equally to the high
overall citation rate discussed in Sect.~\ref{sect:citations}.
The citation rate of the earliest RGS articles is
notably higher.
Although the brightest sources
had quickly been observed with the RGS, later articles
analysing RGS data still receive high citation rates above
three citations per article per year. Also those articles
making use of the OM receive comparably high citation rates.

\subsubsection{Science categories}

\begin{table}
  \caption{Science categories used in Fig.~\ref{fig:categories}}
  \begin{tabular}{lrr}
  \hline\hline
\multicolumn{2}{l}{Science Category \hfill Class codes$^a$} & \#obs$^b$\\
  \hline
$(1)$ Solar System object & 8000-8999 & {\bf 46} \\
$(2)$ Extrasolar Planet/Brown dwarf/Protostar & 1850-1899 & {\bf 65} \\
$(3)$ Stars/WDs &  & {\bf 1062} \\
 \ \ \ \ -- Stars & 1900-2999 & 960 \\
 \ \ \ \ -- White dwarfs & 4000-4999 & 102 \\
$(4)$ Binaries &  & {\bf 1032} \\
 \ \ \ \ -- Cataclysmic Variables & 1600-1699 & 341 \\
 \ \ \ \ -- X-ray binary/QPO & 1000-1599 & 691 \\
$(5)$ Black Holes &  & {\bf 78} \\
 \ \ \ \ -- ULX & 9300-9399 & 61 \\
 \ \ \ \ -- Supermassive black hole & 9400-9499 & 17 \\
$(6)$ Gamma ray source & 1700-1799 & {\bf 161} \\
$(7)$ Pulsar/Neutron star & 1800-1849 & {\bf 478} \\
$(8)$ SNR/Nebulae/Diffuse emission & 3000-3999 & {\bf 1005} \\
$(9)$ Galaxies & 6000-6999 & {\bf 760} \\
$(10)$ Active Galaxies &  & {\bf 1107} \\
 \ \ \ \ -- AGN, Seyferts & 7000-7199 & 869 \\
 \ \ \ \ -- AGN (Liners) & 7400-7499 & 26 \\
 \ \ \ \ -- Radio/IR Galaxies & 7600-7999 & 212 \\
$(11)$ QSO & 7200-7299 & {\bf 452} \\
$(12)$ BL Lac & 7300-7399 & {\bf 182} \\
$(13)$ Radio galaxies & 7500-7599 & {\bf 182} \\
$(14)$ Groups of Galaxies & 5200-5299 & {\bf 182} \\
$(15)$ Clusters of Galaxies & 5000-5099 & {\bf 1152} \\
$(16)$ X-ray background & 5500-5599 & {\bf 117} \\
$(17)$ Supernovae and Hypernovae & 9100-9299 & {\bf 21} \\
$(18)$ Unusual Object & 9000 & {\bf 3} \\
\multicolumn{2}{l}{$(19)$ Cosmology / deep fields \hfill By Target Names$^c$} & {\bf 209} \\
$(20)$ Unidentified & 9999 & {\bf 77} \\
$(21)$ Unclassified &  & {\bf 111} \\
&\hfill Total: &{\bf 8482} \\

  \hline\hline
\label{tab:categories}
  \end{tabular}
$^a$http://heasarc.gsfc.nasa.gov/W3Browse/class\_help.html\\
$^b$Number of observations containing at least one science exposure\\
$^c$All observations with target names containing Marano, Lockman, CDF, AXAF, HDF, COSMOS, MLS, SZE, Groth-Westphal, or Deep Field
\end{table}

Each observation carried out under the GO and ToO programmes
is assigned a science category by the investigator, following the
HEASARC Object Classifications
scheme\footnote{http://heasarc.gsfc.nasa.gov/W3Browse/class\_help.html}.
For calibration observations, the assignments are given by the
instrument team selecting the targets. We combined these codes
into 21 rough groups to study the use
of data taken under different science categories. The grouping
is shown in Table~\ref{tab:categories} where the names of
science categories are listed in the left column, the ranges
of HEASARC class codes in the second column, and the number
of available observations in the XSA in the last column.
We defined an additional category (19) for Cosmology and Deep
Field projects, including
all observations with target names Marano, Lockman, CDF, AXAF, HDF, COSMOS, MLS, SZE, Groth-Westphal, or Deep Field.\\

\begin{figure}[!ht]
\resizebox{\hsize}{!}{\includegraphics{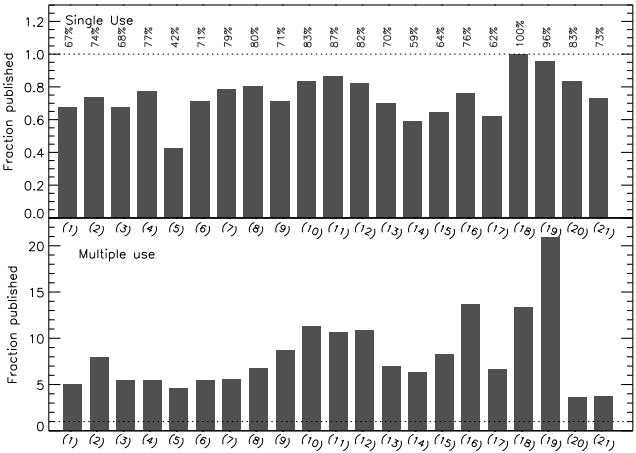}}
\caption{Fractions of observations used in at least one
refereed paper (top panel) for each of the science
categories listed in Table~\ref{tab:categories}.
In the bottom panel, observations used more than once are
counted accordingly multiple times.
The fractions in top and bottom are relative to the total
numbers of observations in each category, given in the
last column of Table~\ref{tab:categories}.
}
\label{fig:categories}
\end{figure}

For each science category listed in Table~\ref{tab:categories},
the fraction of observations that have been used in scientific
articles is shown in the top panel of Fig.~\ref{fig:categories},
with percentages given on top. The numbers in the bottom refer to those in the
first column of Table~\ref{tab:categories}. The recycling rate is
shown in the bottom panel, where multiple use is taken into
account. The histograms show that each observation was used on
average more than five times. The highest recycling rate is in the
Cosmology/Deep field category.

\section{Summary and conclusions}

XMM-Newton, ESA's second X-ray observatory, was launched in 1999 and since then
has been providing the scientific community with high quality imaging and
spectroscopy. The annual calls for observing proposals are typically
over-subscribed by a factor 6 in observing time. In order to investigate
the scientific productivity of XMM-Newton, we have created a database
currently containing information about refereed papers that directly use
XMM-Newton data and the observations that were used. We analysed 3272
articles published until end of 2012 to provide the results reported in
these papers. This information was obtained from
the SAO/NASA Astrophysics Data System (ADS) and
the XMM-Newton Science Archive (XSA).
The main conclusions of this study are:

\begin{itemize}

\item{} Following launch in 1999, the annual number of XMM-Newton refereed
papers rose to 300 per year by 2003 or 2004 and has remained approximately constant
with no evidence for a decrease in recent years.
This continued high level of scientific productivity is remarkable for a mission that is in its
13th year of operation with essentially the same scientific performance as at launch.
It is worth noting that the continued high over-subscription factor in requested
observing time supports the view that XMM-Newton continues to remain
competitive with other facilities.

\item{}
Approximately 100 scientists publish for the first time as first authors
of refereed XMM-Newton papers each year and about 500 appear as co-authors.
The large and unchanging number of new scientists working with XMM-Newton,
suggests that the X-ray community continues to attract new members who are
able to successfully analyse and publish results from the mission.

\item{}
 On average, each XMM-Newton paper
receives four citations each year. The long-term citation rate, counting
citations given more than a few years after publication, remains at a high
average level of three citations per year for each article, five years after
publication. About two thirds of all citations originate from articles that
are not in the XMM-Newton database, thus originate from research that is not
primarily the analysis of XMM-Newton data. Some 50\% of citing articles do not
mention other major X-ray missions, and both citation rates demonstrate the
broad relevance of XMM-Newton results, beyond the X-ray community itself.

\item{} The impact of data from all three instruments is roughly the same in
terms of average citation rate.

\item{}More than 90\% of XMM-Newton observations taken so far have been used in a scientific publication.
Some 10\% of observations have only been used in
projects using several hundred of observations which are probably surveys. Even calibration observations,
that were not primarily taken for scientific purposes, and random slew
observations have been used at levels of 82\% and 22\%, respectively.
An impressively large number of observations has been used multiple
times, indicating the importance of XMM-Newton data beyond the initial
motivation.

\item{} The median time between GO and the first refereed paper being
published ranges from a few months up to several years with the maximum
around two years. We found evidence for a second peak in the distribution
of elapsed times at 3.25 years, almost exactly one year after the first
peak may reflect the one year proprietary period. ToO observations are
published significantly faster with a median time of one year.

\end{itemize}

\acknowledgements
This project benefitted from resources kindly provided by the
XMM-Newton community support and scientific planning team.
We thank Stefan Immler (NASA), Nuria Fonseca Bonilla, Sara Bertran de
Lis, and Pablo Ramirez Moreta for support during screening of articles.
The project was kindly supported by Steve Snowden (NASA Project scientist).
We thank Lauren Boice for useful discussions.
This research has made use of NASA's Astrophysics Data System Bibliographic Services.

\bibliographystyle{aa}
\bibliography{litbib}

\end{document}